\documentclass[showpacs,aps,prd]{revtex4}
\input epsf

\begin{document}

\title{$(Q\bar{s})^{(*)}(\bar{Q}s)^{(*)}$ molecular states from QCD sum rules:\\ a view on $Y(4140)$}
\author{Jian-Rong Zhang and Ming-Qiu Huang}
\affiliation{Department of Physics, National University of Defense
Technology, Hunan 410073, China}

\begin{abstract}
Masses for the $(Q\bar{s})^{(*)}(\bar{Q}s)^{(*)}$ ($Q=c$ or $b$)
molecular states are systematically computed in the framework of QCD
sum rules. Technically, contributions of the operators up to
dimension six are included in operator product expansion (OPE). The
numerical result $4.13\pm0.10~\mbox{GeV}$ for
$D_{s}^{*}\bar{D}_{s}^{*}$ agrees well with the mass
$4143.0\pm2.9\pm1.2~\mbox{MeV}$ for $Y(4140)$, which supports the
$D_{s}^{*}\bar{D}_{s}^{*}$ molecular configuration for $Y(4140)$.
\end{abstract}
\pacs {11.55.Hx, 12.38.Lg, 12.39.Mk}\maketitle
\section{Introduction}\label{sec1}
Recently, the CDF Collaboration has reported the observation of a
narrow near-threshold structure in the $J/\psi\phi$ mass spectrum in
$B^{+}\rightarrow J/\psi\phi K^{+}$ decays \cite{CDF}, for which the
mass is $4143.0\pm2.9\pm1.2~\mbox{MeV}$ and the width is
$11.7^{+8.3}_{-5.0}\pm3.7~\mbox{MeV}$. This experimental observation
has triggered great interest of many practitioners, and there have
already appeared some theoretical interpretations for this new
resonance, e.g. Refs.
\cite{Xiangliu,Mahajan,TBranz,Xiangliu1,GuiJun}. On the whole,
$Y(4140)$ is apt to be deciphered as the molecular partner of the
charmonium-like state $Y(3930)$ \cite{3930}. Undoubtedly, the
quantitative description of $Y(4140)$'s properties such as mass is
quite needed for well understanding its structure, but it is
difficult to extract information on the hadronic spectrum from the
rather simple Lagrangian of QCD. That's because low energy QCD
involves a regime where it is futile to attempt perturbative
calculations and one has to treat a genuinely strong field in
nonperturbative methods. Whereas, one can resort to QCD sum rules
\cite{svzsum} (for reviews see
\cite{overview,overview1,overview2,overview3} and references
therein), which is a nonperturbative analytic formalism firmly
entrenched in QCD. In fact, some authors \cite{zgwang,Nielsen} have
studied $Y(4140)$ via QCD sum rules soon after its observation,
however, they arrived at different conclusions basing on the
$D_{s}^{*}\bar{D}_{s}^{*}$ molecular picture. On the other hand, it
could not be readily excluded for $D_{s}\bar{D}_{s}$ or
$D_{s}^{*}\bar{D}_{s}$ as possible molecular configuration for
$Y(4140)$ without explicit dynamics calculations. Catalyzed by the
above reasons, we devote to calculate the spectra of the
$(Q\bar{s})^{(*)}(\bar{Q}s)^{(*)}$ molecular states through QCD sum
rules, to see whether $Y(4140)$ can be figured as a molecular state.
In our approach, the masses for $D_{s}\bar{D}_{s}$,
$D_{s}^{*}\bar{D}_{s}$, $D_{s}^{*}\bar{D}_{s}^{*}$,
$B_{s}\bar{B}_{s}$, $B_{s}^{*}\bar{B}_{s}$, and
$B_{s}^{*}\bar{B}_{s}^{*}$ molecular states are gained. In addition,
to improve on the accuracy of QCD sum rule analysis for $Y(4140)$,
the $m_{s}^{2}$ order and $\langle g^{3}G^{3}\rangle$ contributions
are included in OPE side.

The paper is organized as follows. In Sec. \ref{sec2}, QCD sum rules
for the molecular states are introduced, and both the
phenomenological representation and QCD side are derived, followed
by the numerical analysis to extract the hadronic masses in Sec.
\ref{sec3}. Section \ref{sec4} is a brief summary.
\section{$(Q\bar{s})^{(*)}(\bar{Q}s)^{(*)}$ QCD sum rules}\label{sec2}
The QCD sum rule attempts to link the hadron phenomenology with the
interactions of quarks and gluons, which contains three main
ingredients: an approximate description of the correlator in terms
of intermediate states through the dispersion relation, a
description of the same correlator in terms of QCD degrees of
freedom via an OPE, and a procedure for matching these two
descriptions and extracting the parameters that characterize the
hadronic state of interest.

\subsection{the molecular state QCD sum rule}
In the molecular pictures, following forms of currents can be
constructed for $(Q\bar{s})^{(*)}(\bar{Q}s)^{(*)}$ states, with
\begin{eqnarray}
j_{(Q\bar{s})(\bar{Q}s)}&=&(\bar{s}_{a}i\gamma_{5}Q_{a})(\bar{Q}_{b}i\gamma_{5}s_{b}),\nonumber\\
j_{(Q\bar{s})^{*}(\bar{Q}s)^{*}}&=&(\bar{s}_{a}\gamma_{\mu}Q_{a})(\bar{Q}_{b}\gamma^{\mu}s_{b}),\nonumber
\end{eqnarray}
for one type of hadrons, and
\begin{eqnarray}
j^{\mu}_{(Q\bar{s})^{*}(\bar{Q}s)}&=&(\bar{s}_{a}\gamma^{\mu}Q_{a})(\bar{Q}_{b}i\gamma_{5}s_{b}),\nonumber
\end{eqnarray}
for another type, where $a$ and $b$ are color indices.

For the former case, the starting point is the two-point correlator
\begin{eqnarray}\label{correlator}
\Pi(q^{2})=i\int
d^{4}x\mbox{e}^{iq.x}\langle0|T[j(x)j^{+}(0)]|0\rangle.
\end{eqnarray}
The correlator can be phenomenologically expressed as a dispersion
integral over a physical spectral function
\begin{eqnarray}
\Pi(q^{2})=\frac{\lambda^{2}_H}{M_{H}^{2}-q^{2}}+\frac{1}{\pi}\int_{s_{0}}
^{\infty}ds\frac{\mbox{Im}\Pi^{\mbox{phen}}(s)}{s-q^{2}}+\mbox{subtractions},
\end{eqnarray}
where $M_{H}$ denotes the mass of the hadronic resonance, and
$\lambda_{H}$ gives the coupling of the current to the hadron
$\langle0|j|H\rangle=\lambda_{H}$. In the OPE side, the correlator
can be written in terms of a dispersion relation as
\begin{eqnarray}
\Pi(q^{2})=\int_{(2m_{Q}+2m_{s})^{2}}^{\infty}ds\frac{\rho^{\mbox{OPE}}(s)}{s-q^{2}},
\end{eqnarray}
where the spectral density is given by the imaginary part of the
correlator
\begin{eqnarray}
\rho^{\mbox{OPE}}(s)=\frac{1}{\pi}\mbox{Im}\Pi^{\mbox{OPE}}(s).
\end{eqnarray}
After equating the two sides, assuming quark-hadron duality, and
making a Borel transform, the sum rule can be written as
\begin{eqnarray}
\lambda_{H}^{2}e^{-M_{H}^{2}/M^{2}}&=&\int_{(2m_{Q}+2m_{s})^{2}}^{s_{0}}ds\rho^{\mbox{OPE}}(s)e^{-s/M^{2}}.
\end{eqnarray}
To eliminate the hadronic coupling constant $\lambda_H$, one reckons
the ratio of derivative of the sum rule and itself, and then yields
\begin{eqnarray}\label{sum rule}
M_{H}^{2}&=&\int_{(2m_{Q}+2m_{s})^{2}}^{s_{0}}ds\rho^{\mbox{OPE}}s
e^{-s/M^{2}}/
\int_{(2m_{Q}+2m_{s})^{2}}^{s_{0}}ds\rho^{\mbox{OPE}}e^{-s/M^{2}}.
\end{eqnarray}

For the latter case, one starts from the two-point correlator
\begin{eqnarray}
\Pi^{\mu\nu}(q^{2})=i\int
d^{4}x\mbox{e}^{iq.x}\langle0|T[j^{\mu}(x)j^{\nu+}(0)]|0\rangle.
\end{eqnarray}
Lorentz covariance implies that the two-point correlation function
can be generally parameterized as
\begin{eqnarray}
\Pi^{\mu\nu}(q^{2})=(\frac{q^{\mu}q^{\nu}}{q^{2}}-g^{\mu\nu})\Pi^{(1)}(q^{2})+\frac{q^{\mu}q^{\nu}}{q^{2}}\Pi^{(0)}(q^{2}).
\end{eqnarray}
The part of the correlator proportional to $g_{\mu\nu}$ will be
chosen to extract the mass sum rule here. In phenomenology,
$\Pi^{(1)}(q^{2})$ can be expressed as a dispersion integral over a
physical spectral function
\begin{eqnarray}
\Pi^{(1)}(q^{2})=\frac{[\lambda^{(1)}]^{2}}{M_{H}^{2}-q^{2}}+\frac{1}{\pi}\int_{s_{0}}
^{\infty}ds\frac{\mbox{Im}\Pi^{(1)\mbox{phen}}(s)}{s-q^{2}}+\mbox{subtractions},
\end{eqnarray}
where $M_{H}$ denotes the mass of the hadronic resonance. In the OPE
side, $\Pi^{(1)}(q^{2})$ can be written in terms of a dispersion
relation as
\begin{eqnarray}
\Pi^{(1)}(q^{2})=\int_{(2m_{Q}+2m_{s})^{2}}^{\infty}ds\frac{\rho^{\mbox{OPE}}(s)}{s-q^{2}},
\end{eqnarray}
where the spectral density is given by
\begin{eqnarray}
\rho^{\mbox{OPE}}(s)=\frac{1}{\pi}\mbox{Im}\Pi^{\mbox{(1)}}(s).
\end{eqnarray}
After equating the two sides, assuming quark-hadron duality, and
making a Borel transform, the sum rule can be written as
\begin{eqnarray}
[\lambda^{(1)}]^{2}e^{-M_{H}^{2}/M^{2}}&=&\int_{(2m_{Q}+2m_{s})^{2}}^{s_{0}}ds\rho^{\mbox{OPE}}(s)e^{-s/M^{2}}.
\end{eqnarray}
To eliminate the hadronic coupling constant $\lambda^{(1)}$, one
reckons the ratio of derivative of the sum rule and itself, and then
yields
\begin{eqnarray}\label{sum rule 1}
M_{H}^{2}&=&\int_{(2m_{Q}+2m_{s})^{2}}^{s_{0}}ds\rho^{\mbox{OPE}}s
e^{-s/M^{2}}/
\int_{(2m_{Q}+2m_{s})^{2}}^{s_{0}}ds\rho^{\mbox{OPE}}e^{-s/M^{2}}.
\end{eqnarray}

\subsection{spectral densities}
Calculating the OPE side, one works at leading order in $\alpha_{s}$
and considers condensates up to dimension six with the similar
techniques in Refs. \cite{technique,technique1}. The $s$ quark is
dealt as a light one and the diagrams are considered up to order
$m_{s}^{2}$. To keep the heavy-quark mass finite, one uses the
momentum-space expression for the heavy-quark propagator. One
calculates the light-quark part of the correlation function in the
coordinate space, which is then Fourier-transformed to the momentum
space in $D$ dimension. The resulting light-quark part is combined
with the heavy-quark part before it is dimensionally regularized at
$D=4$. For the heavy-quark propagator with two and three gluons
attached, the momentum-space expressions given in Ref.
\cite{reinders} are used. After some tedious calculations, finally
with

\begin{eqnarray}
\rho^{\mbox{OPE}}(s)=\rho^{\mbox{pert}}(s)+\rho^{\langle\bar{s}s\rangle}(s)+\rho^{\langle\bar{s}s\rangle^{2}}(s)+\rho^{\langle
g\bar{s}\sigma\cdot G s\rangle}(s)+\rho^{\langle
g^{2}G^{2}\rangle}(s)+\rho^{\langle g^{3}G^{3}\rangle}(s),\nonumber
\end{eqnarray}

\begin{eqnarray}
\rho^{\mbox{pert}}(s)&=&\frac{3}{2^{11}\pi^{6}}\int_{\alpha_{min}}^{\alpha_{max}}\frac{d\alpha}{\alpha^{3}}\int_{\beta_{min}}^{1-\alpha}\frac{d\beta}{\beta^{3}}(1-\alpha-\beta)[(\alpha+\beta)m_{Q}^{2}-\alpha\beta
s]^{4}\nonumber\\&&{}
-\frac{3}{2^{8}\pi^{6}}m_{Q}m_{s}\int_{\alpha_{min}}^{\alpha_{max}}\frac{d\alpha}{\alpha^{3}}\int_{\beta_{min}}^{1-\alpha}\frac{d\beta}{\beta^{2}}(1-\alpha-\beta)[(\alpha+\beta)m_{Q}^{2}-\alpha\beta
s]^{3}\nonumber\\&&{}
+\frac{3^{2}}{2^{9}\pi^{6}}m_{Q}^{2}m_{s}^{2}\int_{\alpha_{min}}^{\alpha_{max}}\frac{d\alpha}{\alpha^{2}}\int_{\beta_{min}}^{1-\alpha}\frac{d\beta}{\beta^{2}}(1-\alpha-\beta)[(\alpha+\beta)m_{Q}^{2}-\alpha\beta
s]^{2},\nonumber
\end{eqnarray}

\begin{eqnarray}
\rho^{\langle\bar{s}s\rangle}(s)&=&-\frac{3\langle\bar{s}s\rangle}{2^{6}\pi^{4}}m_{Q}\int_{\alpha_{min}}^{\alpha_{max}}\frac{d\alpha}{\alpha^{2}}\int_{\beta_{min}}^{1-\alpha}\frac{d\beta}{\beta}[(\alpha+\beta)m_{Q}^{2}-\alpha\beta
s]^{2}\nonumber\\&&{}
+\frac{3\langle\bar{s}s\rangle}{2^{7}\pi^{4}}m_{s}\int_{\alpha_{min}}^{\alpha_{max}}\frac{d\alpha}{\alpha(1-\alpha)}[m_{Q}^{2}-\alpha(1-\alpha)
s]^{2}\nonumber\\&&{}
+\frac{3\langle\bar{s}s\rangle}{2^{5}\pi^{4}}m_{Q}^{2}m_{s}\int_{\alpha_{min}}^{\alpha_{max}}\frac{d\alpha}{\alpha}\int_{\beta_{min}}^{1-\alpha}\frac{d\beta}{\beta}[(\alpha+\beta)m_{Q}^{2}-\alpha\beta
s]\nonumber\\&&{}
-\frac{3\langle\bar{s}s\rangle}{2^{6}\pi^{4}}m_{Q}m_{s}^{2}\int_{\alpha_{min}}^{\alpha_{max}}\frac{d\alpha}{1-\alpha}[m_{Q}^{2}-\alpha(1-\alpha)
s],\nonumber
\end{eqnarray}

\begin{eqnarray}
\rho^{\langle\bar{s}s\rangle^{2}}(s)&=&\frac{\langle\bar{s}s\rangle^{2}}{2^{4}\pi^{2}}m_{Q}^{2}\sqrt{1-4m_{Q}^{2}/s}\nonumber\\&&{}
-\frac{\langle\bar{s}s\rangle^{2}}{2^{4}\pi^{2}}m_{Q}m_{s}\sqrt{1-4m_{Q}^{2}/s}\nonumber\\&&{}
+\frac{3\langle\bar{s}s\rangle^{2}}{2^{5}\pi^{2}}m_{s}^{2}\int_{\alpha_{min}}^{\alpha_{max}}d\alpha\alpha(1-\alpha),\nonumber
\end{eqnarray}

\begin{eqnarray}
\rho^{\langle g\bar{s}\sigma\cdot G s\rangle}(s)&=&-\frac{3\langle
g\bar{s}\sigma\cdot G
s\rangle}{2^{7}\pi^{4}}m_{Q}\int_{\alpha_{min}}^{\alpha_{max}}\frac{d\alpha}{1-\alpha}[m_{Q}^{2}-\alpha(1-\alpha)
s]\nonumber\\&&{} +\frac{\langle g\bar{s}\sigma\cdot G
s\rangle}{2^{7}\pi^{4}}m_{s}\int_{\alpha_{min}}^{\alpha_{max}}d\alpha[2m_{Q}^{2}-3\alpha(1-\alpha)
s]\nonumber\\&&{} +\frac{3\langle g\bar{s}\sigma\cdot G
s\rangle}{2^{7}\pi^{4}}m_{Q}^{2}m_{s}\sqrt{1-4m_{Q}^{2}/s}\nonumber\\&&{}
-\frac{\langle g\bar{s}\sigma\cdot G
s\rangle}{2^{7}\pi^{4}}m_{Q}m_{s}^{2}\sqrt{1-4m_{Q}^{2}/s},\nonumber
\end{eqnarray}

\begin{eqnarray}
\rho^{\langle g^{2}G^{2}\rangle}(s)&=&\frac{\langle
g^{2}G^{2}\rangle}{2^{10}\pi^{6}}m_{Q}^{2}\int_{\alpha_{min}}^{\alpha_{max}}\frac{d\alpha}{\alpha^{3}}\int_{\beta_{min}}^{1-\alpha}d\beta(1-\alpha-\beta)[(\alpha+\beta)m_{Q}^{2}-\alpha\beta
s]\nonumber\\&&{} -\frac{3\langle
g^{2}G^{2}\rangle}{2^{10}\pi^{6}}m_{Q}m_{s}\int_{\alpha_{min}}^{\alpha_{max}}\frac{d\alpha}{\alpha^{3}}\int_{\beta_{min}}^{1-\alpha}d\beta(1-\alpha-\beta)[(\alpha+\beta)m_{Q}^{2}-\alpha\beta
s]\nonumber\\&&{} -\frac{\langle
g^{2}G^{2}\rangle}{2^{10}\pi^{6}}m_{Q}^{3}m_{s}\int_{\alpha_{min}}^{\alpha_{max}}\frac{d\alpha}{\alpha^{3}}\int_{\beta_{min}}^{1-\alpha}d\beta(\alpha+\beta)(1-\alpha-\beta)\nonumber\\&&{}
+\frac{3\langle
g^{2}G^{2}\rangle}{2^{10}\pi^{6}}m_{Q}^{2}m_{s}^{2}\int_{\alpha_{min}}^{\alpha_{max}}\frac{d\alpha}{\alpha^{2}}\int_{\beta_{min}}^{1-\alpha}d\beta(1-\alpha-\beta),\nonumber
\end{eqnarray}

\begin{eqnarray}
\rho^{\langle g^{3}G^{3}\rangle}(s)&=&\frac{\langle
g^{3}G^{3}\rangle}{2^{12}\pi^{6}}\int_{\alpha_{min}}^{\alpha_{max}}\frac{d\alpha}{\alpha^{3}}\int_{\beta_{min}}^{1-\alpha}d\beta(1-\alpha-\beta)[(\alpha+\beta)m_{Q}^{2}-\alpha\beta
s]\nonumber\\&&{} +\frac{\langle
g^{3}G^{3}\rangle}{2^{11}\pi^{6}}m_{Q}^{2}\int_{\alpha_{min}}^{\alpha_{max}}\frac{d\alpha}{\alpha^{3}}\int_{\beta_{min}}^{1-\alpha}d\beta\beta(1-\alpha-\beta)\nonumber\\&&{}
-\frac{\langle
g^{3}G^{3}\rangle}{2^{12}\pi^{6}}m_{Q}m_{s}\int_{\alpha_{min}}^{\alpha_{max}}\frac{d\alpha}{\alpha^{3}}\int_{\beta_{min}}^{1-\alpha}d\beta(\alpha+6\beta)(1-\alpha-\beta),\nonumber
\end{eqnarray}
for $(Q\bar{s})(\bar{Q}s)$,

\begin{eqnarray}
\rho^{\mbox{OPE}}(s)=-\{\rho^{\mbox{pert}}(s)+\rho^{\langle\bar{s}s\rangle}(s)+\rho^{\langle\bar{s}s\rangle^{2}}(s)+\rho^{\langle
g\bar{s}\sigma\cdot G s\rangle}(s)+\rho^{\langle
g^{2}G^{2}\rangle}(s)+\rho^{\langle
g^{3}G^{3}\rangle}(s)\},\nonumber
\end{eqnarray}

\begin{eqnarray}
\rho^{\mbox{pert}}(s)&=&-\frac{3}{2^{12}\pi^{6}}\int_{\alpha_{min}}^{\alpha_{max}}\frac{d\alpha}{\alpha^{3}}\int_{\beta_{min}}^{1-\alpha}\frac{d\beta}{\beta^{3}}(1-\alpha-\beta)(1+\alpha+\beta)[(\alpha+\beta)m_{Q}^{2}-\alpha\beta
s]^{4}\nonumber\\&&{}
+\frac{3}{2^{10}\pi^{6}}m_{Q}m_{s}\int_{\alpha_{min}}^{\alpha_{max}}\frac{d\alpha}{\alpha^{3}}\int_{\beta_{min}}^{1-\alpha}\frac{d\beta}{\beta^{2}}(1-\alpha-\beta)(3+\alpha+\beta)[(\alpha+\beta)m_{Q}^{2}-\alpha\beta
s]^{3}\nonumber\\&&{}
-\frac{3^{2}}{2^{9}\pi^{6}}m_{Q}^{2}m_{s}^{2}\int_{\alpha_{min}}^{\alpha_{max}}\frac{d\alpha}{\alpha^{2}}\int_{\beta_{min}}^{1-\alpha}\frac{d\beta}{\beta^{2}}(1-\alpha-\beta)[(\alpha+\beta)m_{Q}^{2}-\alpha\beta
s]^{2},\nonumber
\end{eqnarray}

\begin{eqnarray}
\rho^{\langle\bar{s}s\rangle}(s)&=&\frac{3\langle\bar{s}s\rangle}{2^{7}\pi^{4}}m_{Q}\int_{\alpha_{min}}^{\alpha_{max}}\frac{d\alpha}{\alpha^{2}}\int_{\beta_{min}}^{1-\alpha}\frac{d\beta}{\beta}(1+\alpha+\beta)[(\alpha+\beta)m_{Q}^{2}-\alpha\beta
s]^{2}\nonumber\\&&{}
+\frac{3\langle\bar{s}s\rangle}{2^{7}\pi^{4}}m_{s}\int_{\alpha_{min}}^{\alpha_{max}}\frac{d\alpha}{\alpha}\int_{\beta_{min}}^{1-\alpha}\frac{d\beta}{\beta}[(\alpha+\beta)m_{Q}^{2}-\alpha\beta
s]^{2}\nonumber\\&&{}
-\frac{3\langle\bar{s}s\rangle}{2^{7}\pi^{4}}m_{s}\int_{\alpha_{min}}^{\alpha_{max}}\frac{d\alpha}{\alpha(1-\alpha)}[m_{Q}^{2}-\alpha(1-\alpha)
s]^{2}\nonumber\\&&{}
-\frac{3\langle\bar{s}s\rangle}{2^{5}\pi^{4}}m_{Q}^{2}m_{s}\int_{\alpha_{min}}^{\alpha_{max}}\frac{d\alpha}{\alpha}\int_{\beta_{min}}^{1-\alpha}\frac{d\beta}{\beta}[(\alpha+\beta)m_{Q}^{2}-\alpha\beta
s]\nonumber\\&&{}
+\frac{3\langle\bar{s}s\rangle}{2^{6}\pi^{4}}m_{Q}m_{s}^{2}\int_{\alpha_{min}}^{\alpha_{max}}\frac{d\alpha}{1-\alpha}[m_{Q}^{2}-\alpha(1-\alpha)
s]\nonumber\\&&{}
-\frac{3\langle\bar{s}s\rangle}{2^{7}\pi^{4}}m_{Q}m_{s}^{2}\int_{\alpha_{min}}^{\alpha_{max}}d\alpha\int_{\beta_{min}}^{1-\alpha}\frac{d\beta}{\beta}[(\alpha+\beta)m_{Q}^{2}-\alpha\beta
s],\nonumber
\end{eqnarray}

\begin{eqnarray}
\rho^{\langle\bar{s}s\rangle^{2}}(s)&=&-\frac{\langle\bar{s}s\rangle^{2}}{2^{4}\pi^{2}}m_{Q}^{2}\sqrt{1-4m_{Q}^{2}/s}\nonumber\\&&{}
+\frac{3\langle\bar{s}s\rangle^{2}}{2^{6}\pi^{2}}m_{Q}m_{s}\sqrt{1-4m_{Q}^{2}/s}\nonumber\\&&{}
-\frac{3\langle\bar{s}s\rangle^{2}}{2^{6}\pi^{2}}m_{s}^{2}\int_{\alpha_{min}}^{\alpha_{max}}d\alpha\alpha(1-\alpha),\nonumber
\end{eqnarray}

\begin{eqnarray}
\rho^{\langle g\bar{s}\sigma\cdot G s\rangle}(s)&=&-\frac{3\langle
g\bar{s}\sigma\cdot G
s\rangle}{2^{8}\pi^{4}}m_{Q}\int_{\alpha_{min}}^{\alpha_{max}}d\alpha\int_{\beta_{min}}^{1-\alpha}\frac{d\beta}{\beta}[(\alpha+\beta)m_{Q}^{2}-\alpha\beta
s]\nonumber\\&&{}+\frac{3\langle g\bar{s}\sigma\cdot G
s\rangle}{2^{7}\pi^{4}}m_{Q}\int_{\alpha_{min}}^{\alpha_{max}}\frac{d\alpha}{1-\alpha}[m_{Q}^{2}-\alpha(1-\alpha)
s]\nonumber\\&&{}-\frac{\langle g\bar{s}\sigma\cdot G
s\rangle}{2^{7}\pi^{4}}m_{s}\int_{\alpha_{min}}^{\alpha_{max}}d\alpha[m_{Q}^{2}-2\alpha(1-\alpha)
s]\nonumber\\&&{} -\frac{3\langle g\bar{s}\sigma\cdot G
s\rangle}{2^{7}\pi^{4}}m_{Q}^{2}m_{s}\sqrt{1-4m_{Q}^{2}/s}\nonumber\\&&{}
+\frac{3\langle g\bar{s}\sigma\cdot G
s\rangle}{2^{9}\pi^{4}}m_{Q}m_{s}^{2}\sqrt{1-4m_{Q}^{2}/s},\nonumber
\end{eqnarray}

\begin{eqnarray}
\rho^{\langle g^{2}G^{2}\rangle}(s)&=&-\frac{\langle
g^{2}G^{2}\rangle}{2^{11}\pi^{6}}m_{Q}^{2}\int_{\alpha_{min}}^{\alpha_{max}}\frac{d\alpha}{\alpha^{3}}\int_{\beta_{min}}^{1-\alpha}d\beta(1-\alpha-\beta)(1+\alpha+\beta)[(\alpha+\beta)m_{Q}^{2}-\alpha\beta
s]\nonumber\\&&{} +\frac{3\langle
g^{2}G^{2}\rangle}{2^{12}\pi^{6}}m_{Q}m_{s}\int_{\alpha_{min}}^{\alpha_{max}}\frac{d\alpha}{\alpha^{3}}\int_{\beta_{min}}^{1-\alpha}d\beta(1-\alpha-\beta)(3+\alpha+\beta)[(\alpha+\beta)m_{Q}^{2}-\alpha\beta
s]\nonumber\\&&{} +\frac{\langle
g^{2}G^{2}\rangle}{2^{12}\pi^{6}}m_{Q}^{3}m_{s}\int_{\alpha_{min}}^{\alpha_{max}}\frac{d\alpha}{\alpha^{3}}\int_{\beta_{min}}^{1-\alpha}d\beta(\alpha+\beta)(1-\alpha-\beta)(3+\alpha+\beta)\nonumber\\&&{}
-\frac{3\langle
g^{2}G^{2}\rangle}{2^{10}\pi^{6}}m_{Q}^{2}m_{s}^{2}\int_{\alpha_{min}}^{\alpha_{max}}\frac{d\alpha}{\alpha^{2}}\int_{\beta_{min}}^{1-\alpha}d\beta(1-\alpha-\beta),\nonumber
\end{eqnarray}

\begin{eqnarray}
\rho^{\langle g^{3}G^{3}\rangle}(s)&=&-\frac{\langle
g^{3}G^{3}\rangle}{2^{13}\pi^{6}}\int_{\alpha_{min}}^{\alpha_{max}}\frac{d\alpha}{\alpha^{3}}\int_{\beta_{min}}^{1-\alpha}d\beta(1-\alpha-\beta)(1+\alpha+\beta)[(\alpha+\beta)m_{Q}^{2}-\alpha\beta
s]\nonumber\\&&{} -\frac{\langle
g^{3}G^{3}\rangle}{2^{12}\pi^{6}}m_{Q}^{2}\int_{\alpha_{min}}^{\alpha_{max}}\frac{d\alpha}{\alpha^{3}}\int_{\beta_{min}}^{1-\alpha}d\beta\beta(1-\alpha-\beta)(1+\alpha+\beta)\nonumber\\&&{}
+\frac{\langle
g^{3}G^{3}\rangle}{2^{14}\pi^{6}}m_{Q}m_{s}\int_{\alpha_{min}}^{\alpha_{max}}\frac{d\alpha}{\alpha^{3}}\int_{\beta_{min}}^{1-\alpha}d\beta(\alpha+6\beta)(1-\alpha-\beta)(3+\alpha+\beta),\nonumber
\end{eqnarray}
for $(Q\bar{s})^{*}(\bar{Q}s)$, and

\begin{eqnarray}
\rho^{\mbox{OPE}}(s)=\rho^{\mbox{pert}}(s)+\rho^{\langle\bar{s}s\rangle}(s)+\rho^{\langle\bar{s}s\rangle^{2}}(s)+\rho^{\langle
g\bar{s}\sigma\cdot G s\rangle}(s)+\rho^{\langle
g^{2}G^{2}\rangle}(s)+\rho^{\langle g^{3}G^{3}\rangle}(s),\nonumber
\end{eqnarray}

\begin{eqnarray}
\rho^{\mbox{pert}}(s)&=&\frac{3}{2^{9}\pi^{6}}\int_{\alpha_{min}}^{\alpha_{max}}\frac{d\alpha}{\alpha^{3}}\int_{\beta_{min}}^{1-\alpha}\frac{d\beta}{\beta^{3}}(1-\alpha-\beta)[(\alpha+\beta)m_{Q}^{2}-\alpha\beta
s]^{4}\nonumber\\&&{}
-\frac{3}{2^{7}\pi^{6}}m_{Q}m_{s}\int_{\alpha_{min}}^{\alpha_{max}}\frac{d\alpha}{\alpha^{3}}\int_{\beta_{min}}^{1-\alpha}\frac{d\beta}{\beta^{2}}(1-\alpha-\beta)[(\alpha+\beta)m_{Q}^{2}-\alpha\beta
s]^{3}\nonumber\\&&{}
+\frac{3^{2}}{2^{7}\pi^{6}}m_{Q}^{2}m_{s}^{2}\int_{\alpha_{min}}^{\alpha_{max}}\frac{d\alpha}{\alpha^{2}}\int_{\beta_{min}}^{1-\alpha}\frac{d\beta}{\beta^{2}}(1-\alpha-\beta)[(\alpha+\beta)m_{Q}^{2}-\alpha\beta
s]^{2},\nonumber
\end{eqnarray}

\begin{eqnarray}
\rho^{\langle\bar{s}s\rangle}(s)&=&-\frac{3\langle\bar{s}s\rangle}{2^{5}\pi^{4}}m_{Q}\int_{\alpha_{min}}^{\alpha_{max}}\frac{d\alpha}{\alpha^{2}}\int_{\beta_{min}}^{1-\alpha}\frac{d\beta}{\beta}[(\alpha+\beta)m_{Q}^{2}-\alpha\beta
s]^{2}\nonumber\\&&{}
+\frac{3\langle\bar{s}s\rangle}{2^{5}\pi^{4}}m_{s}\int_{\alpha_{min}}^{\alpha_{max}}\frac{d\alpha}{\alpha(1-\alpha)}[m_{Q}^{2}-\alpha(1-\alpha)
s]^{2}\nonumber\\&&{}
+\frac{3\langle\bar{s}s\rangle}{2^{3}\pi^{4}}m_{Q}^{2}m_{s}\int_{\alpha_{min}}^{\alpha_{max}}\frac{d\alpha}{\alpha}\int_{\beta_{min}}^{1-\alpha}\frac{d\beta}{\beta}[(\alpha+\beta)m_{Q}^{2}-\alpha\beta
s]\nonumber\\&&{}
-\frac{3\langle\bar{s}s\rangle}{2^{5}\pi^{4}}m_{Q}m_{s}^{2}\int_{\alpha_{min}}^{\alpha_{max}}\frac{d\alpha}{1-\alpha}[m_{Q}^{2}-\alpha(1-\alpha)
s],\nonumber
\end{eqnarray}

\begin{eqnarray}
\rho^{\langle\bar{s}s\rangle^{2}}(s)&=&\frac{\langle\bar{s}s\rangle^{2}}{2^{2}\pi^{2}}m_{Q}^{2}\sqrt{1-4m_{Q}^{2}/s}\nonumber\\&&{}
-\frac{\langle\bar{s}s\rangle^{2}}{2^{3}\pi^{2}}m_{Q}m_{s}\sqrt{1-4m_{Q}^{2}/s}\nonumber\\&&{}
+\frac{3\langle\bar{s}s\rangle^{2}}{2^{3}\pi^{2}}m_{s}^{2}\int_{\alpha_{min}}^{\alpha_{max}}d\alpha\alpha(1-\alpha),\nonumber
\end{eqnarray}

\begin{eqnarray}
\rho^{\langle g\bar{s}\sigma\cdot G s\rangle}(s)&=&-\frac{3\langle
g\bar{s}\sigma\cdot G
s\rangle}{2^{6}\pi^{4}}m_{Q}\int_{\alpha_{min}}^{\alpha_{max}}\frac{d\alpha}{1-\alpha}[m_{Q}^{2}-\alpha(1-\alpha)
s]\nonumber\\&&{} +\frac{\langle g\bar{s}\sigma\cdot G
s\rangle}{2^{5}\pi^{4}}m_{s}\int_{\alpha_{min}}^{\alpha_{max}}d\alpha[2m_{Q}^{2}-3\alpha(1-\alpha)
s]\nonumber\\&&{} +\frac{3\langle g\bar{s}\sigma\cdot G
s\rangle}{2^{5}\pi^{4}}m_{Q}^{2}m_{s}\sqrt{1-4m_{Q}^{2}/s}\nonumber\\&&{}
-\frac{\langle g\bar{s}\sigma\cdot G
s\rangle}{2^{6}\pi^{4}}m_{Q}m_{s}^{2}\sqrt{1-4m_{Q}^{2}/s},\nonumber
\end{eqnarray}

\begin{eqnarray}
\rho^{\langle g^{2}G^{2}\rangle}(s)&=&\frac{\langle
g^{2}G^{2}\rangle}{2^{8}\pi^{6}}m_{Q}^{2}\int_{\alpha_{min}}^{\alpha_{max}}\frac{d\alpha}{\alpha^{3}}\int_{\beta_{min}}^{1-\alpha}d\beta(1-\alpha-\beta)[(\alpha+\beta)m_{Q}^{2}-\alpha\beta
s]\nonumber\\&&{} -\frac{3\langle
g^{2}G^{2}\rangle}{2^{9}\pi^{6}}m_{Q}m_{s}\int_{\alpha_{min}}^{\alpha_{max}}\frac{d\alpha}{\alpha^{3}}\int_{\beta_{min}}^{1-\alpha}d\beta(1-\alpha-\beta)[(\alpha+\beta)m_{Q}^{2}-\alpha\beta
s]\nonumber\\&&{} -\frac{\langle
g^{2}G^{2}\rangle}{2^{9}\pi^{6}}m_{Q}^{3}m_{s}\int_{\alpha_{min}}^{\alpha_{max}}\frac{d\alpha}{\alpha^{3}}\int_{\beta_{min}}^{1-\alpha}d\beta(\alpha+\beta)(1-\alpha-\beta)\nonumber\\&&{}
+\frac{3\langle
g^{2}G^{2}\rangle}{2^{8}\pi^{6}}m_{Q}^{2}m_{s}^{2}\int_{\alpha_{min}}^{\alpha_{max}}\frac{d\alpha}{\alpha^{2}}\int_{\beta_{min}}^{1-\alpha}d\beta(1-\alpha-\beta),\nonumber
\end{eqnarray}

\begin{eqnarray}
\rho^{\langle g^{3}G^{3}\rangle}(s)&=&\frac{\langle
g^{3}G^{3}\rangle}{2^{10}\pi^{6}}\int_{\alpha_{min}}^{\alpha_{max}}\frac{d\alpha}{\alpha^{3}}\int_{\beta_{min}}^{1-\alpha}d\beta(1-\alpha-\beta)[(\alpha+\beta)m_{Q}^{2}-\alpha\beta
s]\nonumber\\&&{} +\frac{\langle
g^{3}G^{3}\rangle}{2^{9}\pi^{6}}m_{Q}^{2}\int_{\alpha_{min}}^{\alpha_{max}}\frac{d\alpha}{\alpha^{3}}\int_{\beta_{min}}^{1-\alpha}d\beta\beta(1-\alpha-\beta)\nonumber\\&&{}
-\frac{\langle
g^{3}G^{3}\rangle}{2^{11}\pi^{6}}m_{Q}m_{s}\int_{\alpha_{min}}^{\alpha_{max}}\frac{d\alpha}{\alpha^{3}}\int_{\beta_{min}}^{1-\alpha}d\beta(\alpha+6\beta)(1-\alpha-\beta),\nonumber
\end{eqnarray}
for $(Q\bar{s})^{*}(\bar{Q}s)^{*}$. The integration limits are given
by $\alpha_{min}=(1-\sqrt{1-4m_{Q}^{2}/s})/2$,
$\alpha_{max}=(1+\sqrt{1-4m_{Q}^{2}/s})/2$, and $\beta_{min}=\alpha
m_{Q}^{2}/(s\alpha-m_{Q}^{2})$.
\section{Numerical analysis}\label{sec3}
In this part, the sum rules (\ref{sum rule}) and (\ref{sum rule 1})
will be numerically analyzed. The input values are taken as
$m_{c}=1.23~\mbox{GeV}$, $m_{b}=4.20~\mbox{GeV}$,
$m_{s}=0.13~\mbox{GeV}$,
$\langle\bar{q}q\rangle=-(0.23)^{3}~\mbox{GeV}^{3}$,
$\langle\bar{s}s\rangle=0.8~\langle\bar{q}q\rangle$, $\langle
g\bar{s}\sigma\cdot G s\rangle=m_{0}^{2}~\langle\bar{s}s\rangle$,
$m_{0}^{2}=0.8~\mbox{GeV}^{2}$, $\langle
g^{2}G^{2}\rangle=0.88~\mbox{GeV}^{4}$, and $\langle
g^{3}G^{3}\rangle=0.045~\mbox{GeV}^{6}$. Complying with the standard
procedure of sum rule analysis, the threshold $s_{0}$ and Borel
parameter $M^{2}$ are varied to find the optimal stability window,
in which the perturbative contribution should be larger than the
condensate contributions while the pole contribution larger than
continuum contribution. Thus, the regions of $s_{0}$ and $M^{2}$ are
taken as $\sqrt{s_0}=4.3\sim4.5~\mbox{GeV}$,
$M^{2}=3.5\sim4.5~\mbox{GeV}^{2}$ for $D_{s}\bar{D}_{s}$,
$\sqrt{s_0}=4.5\sim4.7~\mbox{GeV}$,
$M^{2}=3.5\sim4.5~\mbox{GeV}^{2}$ for $D_{s}^{*}\bar{D}_{s}$,
$\sqrt{s_0}=4.6\sim4.8~\mbox{GeV}$,
$M^{2}=3.5\sim4.5~\mbox{GeV}^{2}$ for $D_{s}^{*}\bar{D}_{s}^{*}$,
$\sqrt{s_0}=11.1\sim11.3~\mbox{GeV}$,
$M^{2}=9.5\sim11.0~\mbox{GeV}^{2}$ for $B_{s}\bar{B}_{s}$,
$\sqrt{s_0}=11.1\sim11.3~\mbox{GeV}$,
$M^{2}=9.5\sim11.0~\mbox{GeV}^{2}$ for $B_{s}^{*}\bar{B}_{s}$, and
$\sqrt{s_0}=11.2\sim11.4~\mbox{GeV}$,
$M^{2}=9.5\sim11.0~\mbox{GeV}^{2}$ for $B_{s}^{*}\bar{B}_{s}^{*}$,
respectively. The corresponding Borel curves are exhibited in Figs.
1-3. Ultimately, we obtain the mass values: $3.91\pm0.10~\mbox{GeV}$
for $D_{s}\bar{D}_{s}$, $4.01\pm0.10~\mbox{GeV}$ for
$D_{s}^{*}\bar{D}_{s}$, $4.13\pm0.10~\mbox{GeV}$ for
$D_{s}^{*}\bar{D}_{s}^{*}$, $10.70\pm0.10~\mbox{GeV}$ for
$B_{s}\bar{B}_{s}$, $10.71\pm0.11~\mbox{GeV}$ for
$B_{s}^{*}\bar{B}_{s}$, and $10.80\pm0.10~\mbox{GeV}$ for
$B_{s}^{*}\bar{B}_{s}^{*}$. It is worth noting that uncertainty in
our results are merely owing to the sum rule windows (variation of
the threshold $s_{0}$ and Borel parameter $M^{2}$), not involving
the ones from the variation of quark masses and QCD parameters.

\begin{figure}[htb!]
\centerline{\epsfysize=5truecm
\epsfbox{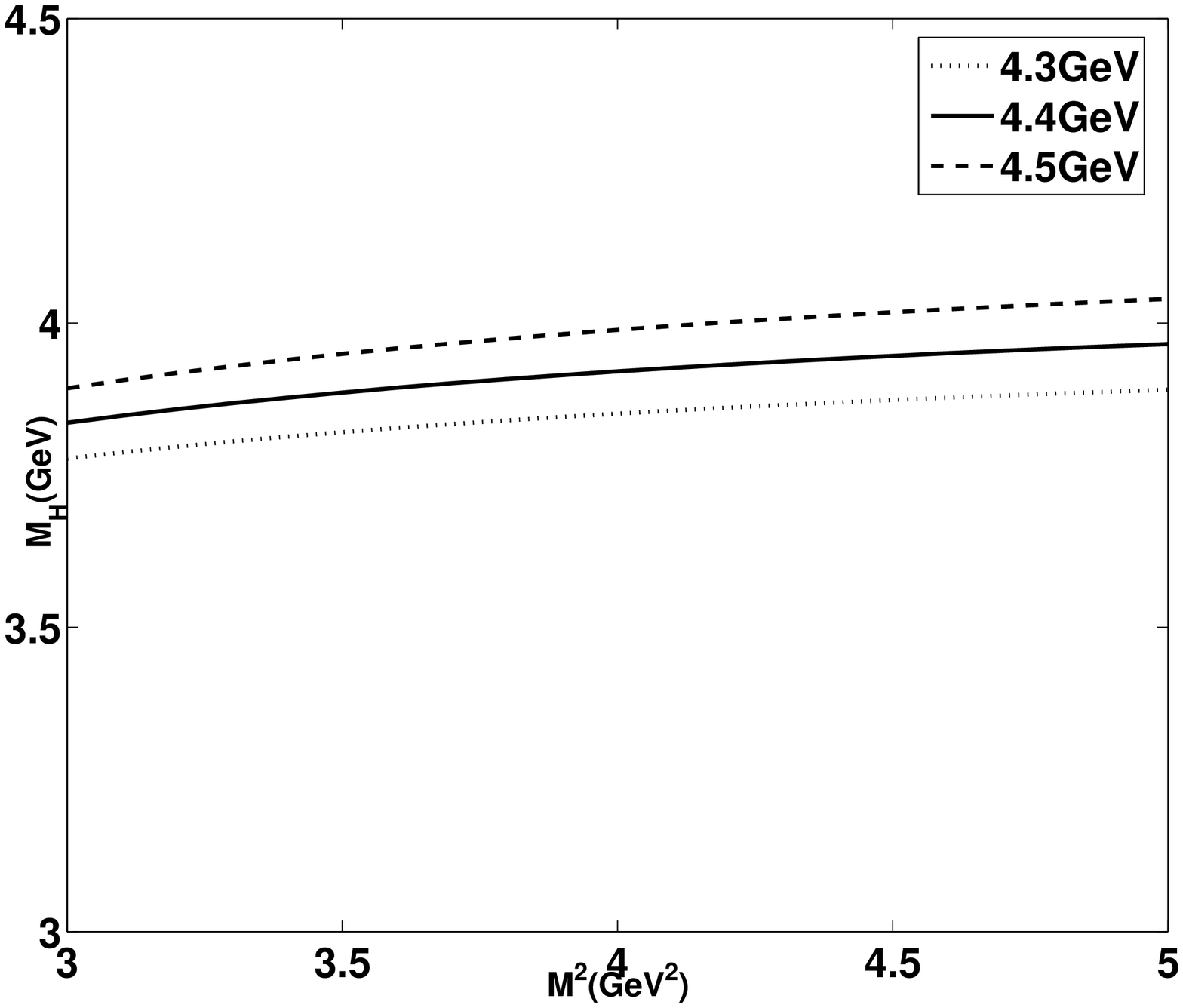}\epsfysize=5truecm\epsfbox{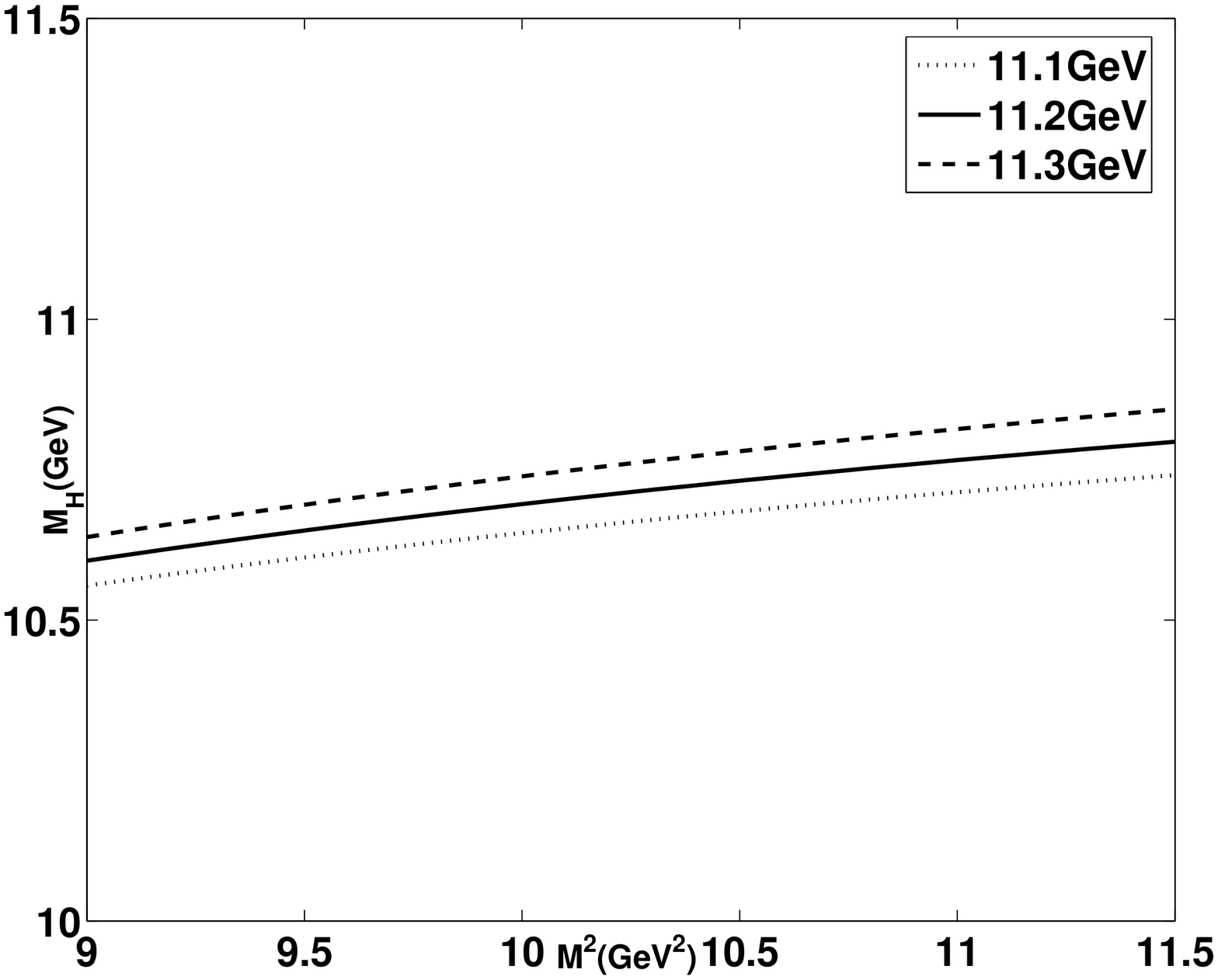}}\caption{The
dependence on $M^2$ for the masses of $D_{s}\bar{D}_{s}$ and
$B_{s}\bar{B}_{s}$ from sum rule (\ref{sum rule}). The continuum
thresholds are taken as $\sqrt{s_0}=4.3\sim4.5~\mbox{GeV}$ and
$\sqrt{s_0}=11.1\sim11.3~\mbox{GeV}$, respectively.} \label{fig:1}
\end{figure}

\begin{figure}
\centerline{\epsfysize=5truecm
\epsfbox{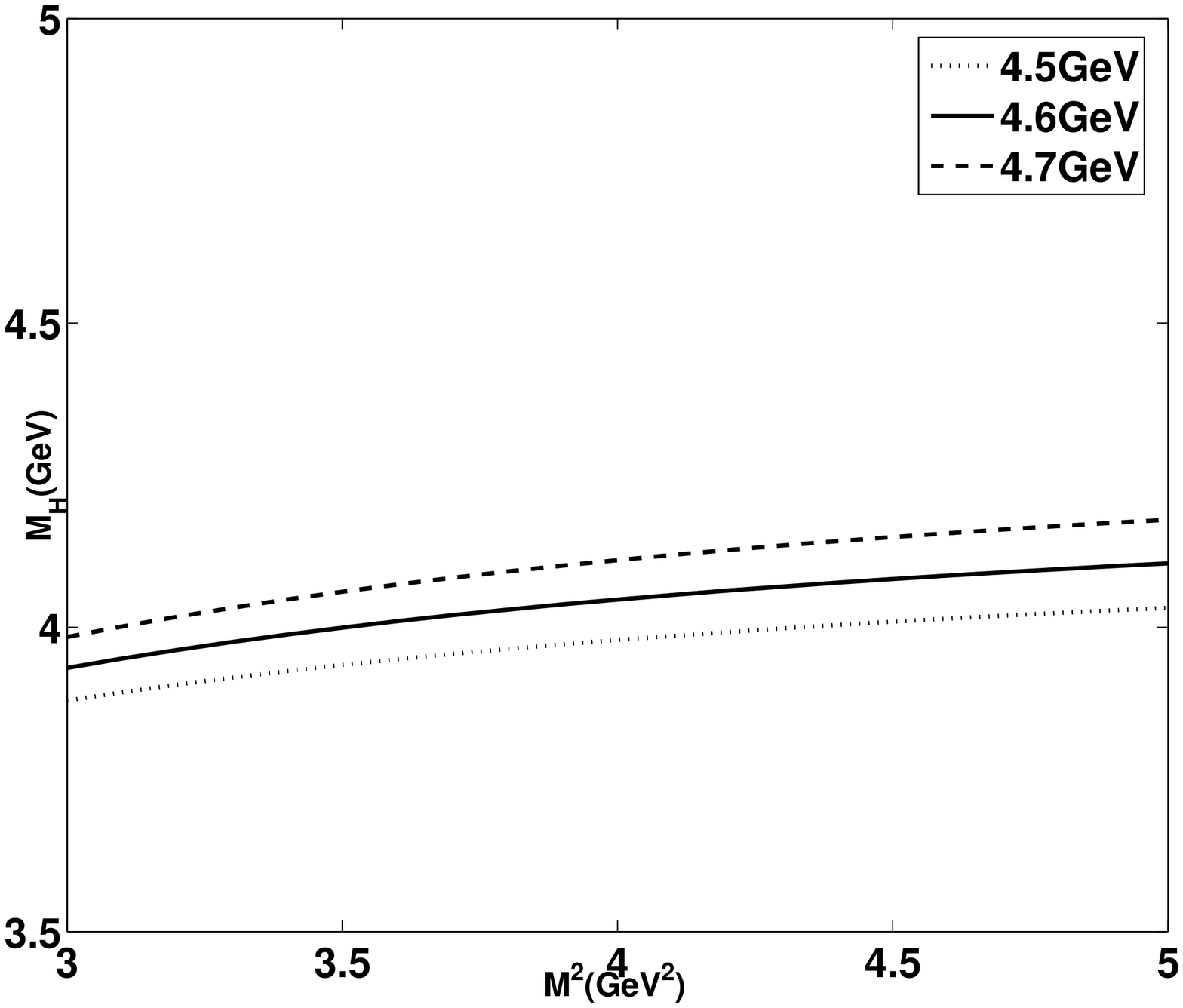}\epsfysize=5truecm\epsfbox{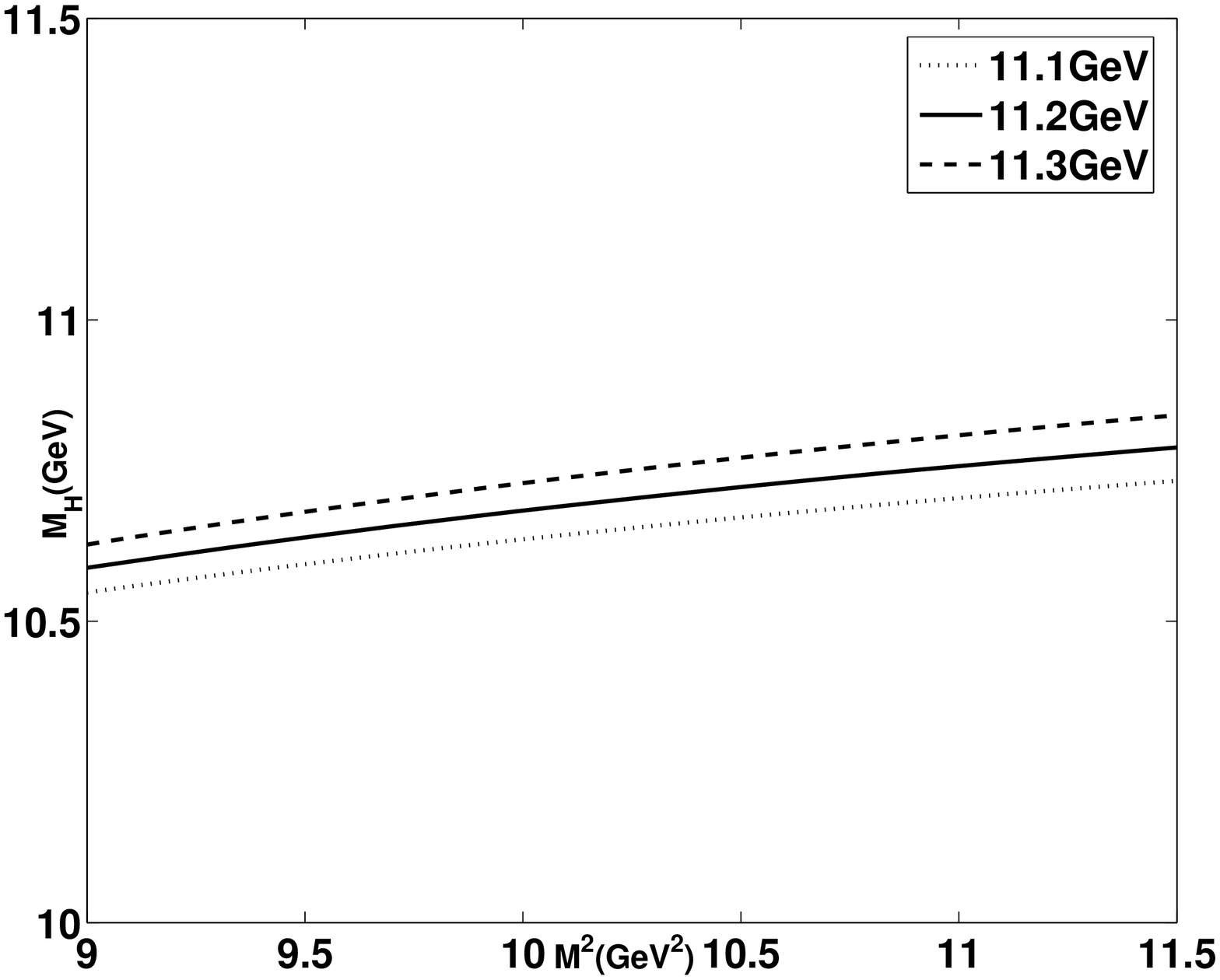}}\caption{The
dependence on $M^2$ for the masses of $D_{s}^{*}\bar{D}_{s}$ and
$B_{s}^{*}\bar{B}_{s}$ from sum rule (\ref{sum rule 1}). The
continuum thresholds are taken as $\sqrt{s_0}=4.5\sim4.7~\mbox{GeV}$
and $\sqrt{s_0}=11.1\sim11.3~\mbox{GeV}$, respectively.}
\label{fig:2}
\end{figure}

\begin{figure}
\centerline{\epsfysize=5truecm
\epsfbox{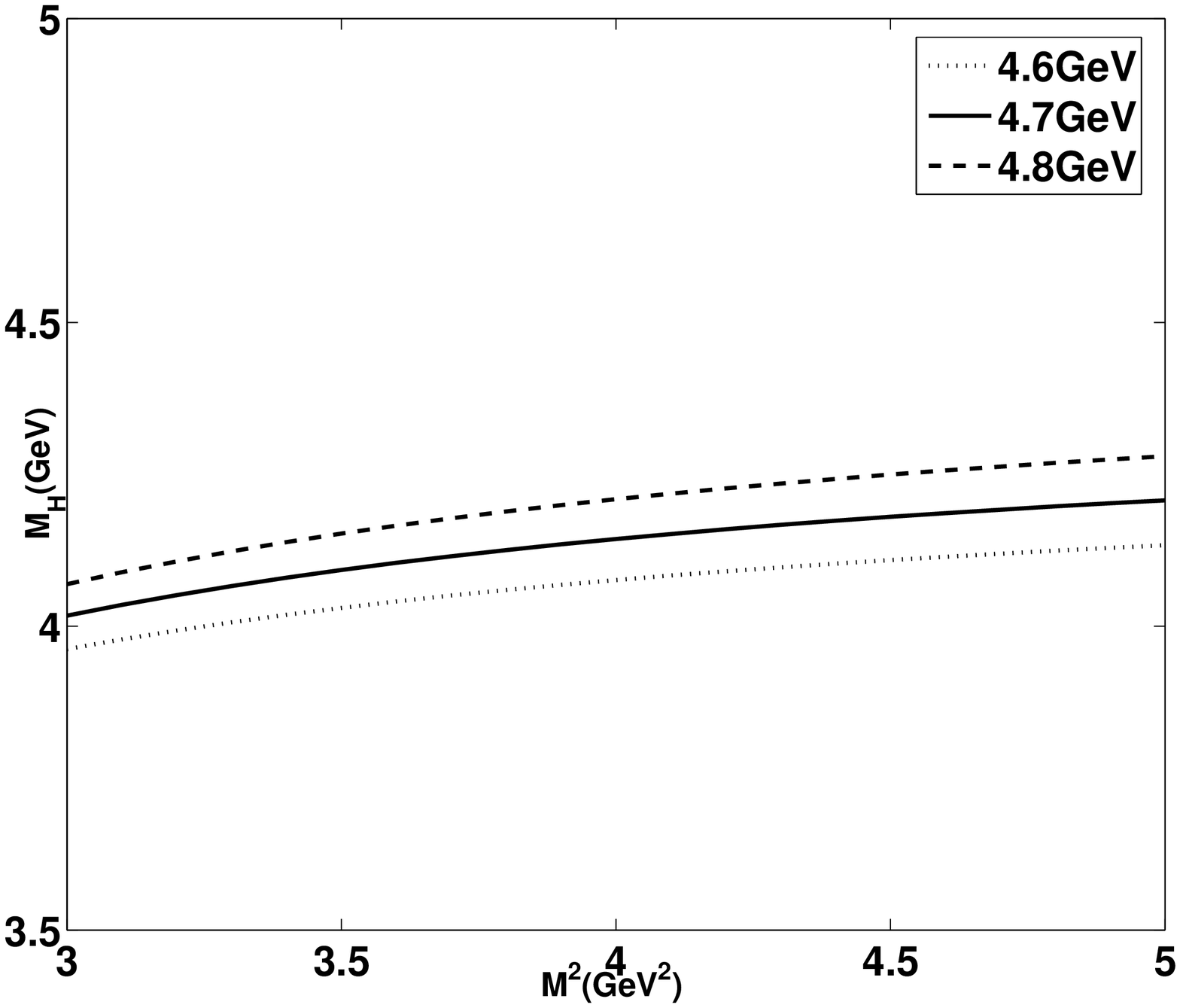}\epsfysize=5truecm\epsfbox{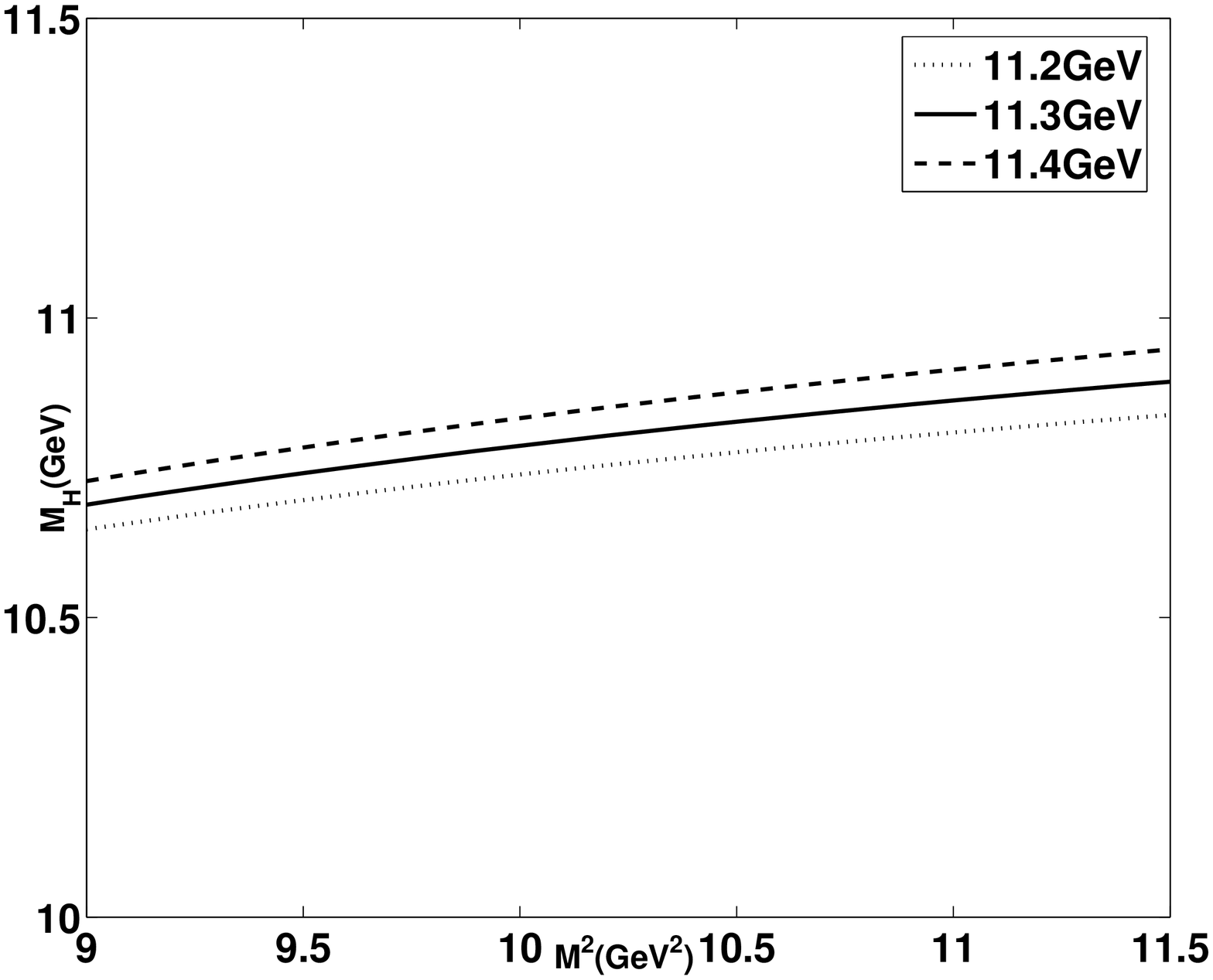}}\caption{The
dependence on $M^2$ for the masses of $D_{s}^{*}\bar{D}_{s}^{*}$ and
$B_{s}^{*}\bar{B}_{s}^{*}$ from sum rule (\ref{sum rule}). The
continuum thresholds are taken as $\sqrt{s_0}=4.6\sim4.8~\mbox{GeV}$
and $\sqrt{s_0}=11.2\sim11.4~\mbox{GeV}$, respectively.}
\label{fig:3}
\end{figure}

\section{Summary}\label{sec4}
In summary, the QCD sum rules have been employed to compute the
masses of $(Q\bar{s})^{(*)}(\bar{Q}s)^{(*)}$, including the
contributions of the operators up to dimension six in OPE.  For the
charmed molecular states, we have got
$M_{D_{s}\bar{D}_{s}}=3.91\pm0.10~\mbox{GeV}$,
$M_{D_{s}^{*}\bar{D}_{s}}=4.01\pm0.10~\mbox{GeV}$, and
$M_{D_{s}^{*}\bar{D}_{s}^{*}}=4.13\pm0.10~\mbox{GeV}$. The numerical
values for $D_{s}\bar{D}_{s}$ and $D_{s}^{*}\bar{D}_{s}$ are lower
than the mass of $Y(4140)$, $4143.0\pm2.9\pm1.2~\mbox{MeV}$,
whereas, the one for $D_{s}^{*}\bar{D}_{s}^{*}$ is well compatible
with the experimental data, which supports the
$D_{s}^{*}\bar{D}_{s}^{*}$ configuration for $Y(4140)$.
Additionally, we have extracted
$M_{B_{s}\bar{B}_{s}}=10.70\pm0.10~\mbox{GeV}$,
$M_{B_{s}^{*}\bar{B}_{s}}=10.71\pm0.11~\mbox{GeV}$, and
$M_{B_{s}^{*}\bar{B}_{s}^{*}}=10.80\pm0.10~\mbox{GeV}$ for the
bottom molecular states. Altogether, all these theoretical results
are looking forward to further experimental identification.

\begin{acknowledgments}
J. R. Zhang is very indebted to Ming Zhong for helpful discussions.
This work was supported in part by the National Natural Science
Foundation of China under Contract No.10675167.
\end{acknowledgments}

\end{document}